\documentclass[aps,prc,reprint,amsmath,amssymb,showpacs,superscriptaddress]{revtex4-1}
\usepackage{CJK}
\usepackage{graphicx}
\usepackage{dcolumn}
\usepackage{bm}

\usepackage{color}
\usepackage{hyperref}

\begin{document}
\begin{CJK*}{UTF8}{} 


\title{Relativistic orbital-free kinetic energy density functional for one-particle nuclear systems}

\author{X. H. Wu}
\email{wuxinhui@fzu.edu.cn}
\affiliation{Department of Physics, Fuzhou University, Fuzhou 350108, Fujian, China}
\affiliation{State Key Laboratory of Nuclear Physics and Technology, School of Physics, Peking University, Beijing 100871, China}

\author{Z. X. Ren}
\affiliation{Institute for Advanced Simulation, Forschungszentrum J\"{u}lich, D-52425 J\"{u}lich, Germany}
\affiliation{Helmholtz-Institut f\"{u}r Strahlen-und Kernphysik and Bethe Center for Theoretical Physics, Universit\"{a}t Bonn, D-53115 Bonn, Germany}

\author{H. Z. Liang}
\affiliation{Department of Physics, Graduate School of Science, The University of Tokyo, Tokyo 113-0033, Japan}
\affiliation{Quark Nuclear Science Institute, The University of Tokyo, Tokyo 113-0033, Japan}
\affiliation{RIKEN Center for Interdisciplinary Theoretical and Mathematical Sciences (iTHEMS), Wako 351-0198, Japan}

\author{P. W. Zhao}
\email{pwzhao@pku.edu.cn}
\affiliation{State Key Laboratory of Nuclear Physics and Technology, School of Physics, Peking University, Beijing 100871, China}


\begin{abstract}
  This letter aims to derive the exact relativistic orbital-free kinetic energy density functional for one-particle nuclear systems in one-dimensional case.
  The kinetic energy is expressed as a functional of both vector and scalar densities.
  The functional derivatives of the kinetic energy density functional are also derived.
  Both the kinetic energy density functional and its functional derivatives are validated to be correct.
  This serves as a foundation for further exploration of more general relativistic orbital-free  kinetic energy density functionals.
\end{abstract}


\maketitle

\end{CJK*}



\section{Introduction}

Research on quantum many-body systems is essential across a wide range of scientific fields.
Directly solving the quantum many-body Schr{\"o}dinger equation exhibits exponential computational complexity with increasing number of particles.
Density functional theory (DFT), based on the Hohenberg-Kohn theorem~\cite{Hohenberg1964Phys.Rev.},  provides fully quantum solutions at a fraction of the cost of directly solving the Schr{\"o}dinger equation by mapping the coupled many-body problem to a single-particle problem.
DFT has been widely applied in nuclear physics, predominantly within the Kohn-Sham (KS) scheme~\cite{Kohn1965Phys.Rev.}, which introduces auxiliary one-body orbitals to compute the kinetic energy.

On the other hand, in the scheme of orbital-free DFT (OF-DFT)~\cite{Levy1984PRA}, one aims to express the energy solely as a functional of the density.
The motivation for developing OF-DFT comes from two fronts.
On the theoretical front, OF-DFT returns to the roots of the Hohenberg-Kohn theorem and puts the density back into centrality in the DFT study, making it a more fundamental framework than Kohn-Sham DFT.
It can facilitate a deeper understanding of the Hohenberg-Kohn theorem and has long been an ideal pursuit for many DFT researchers.
On the practical front, OF-DFT can be much more efficient than the Kohn-Sham DFT.
OF-DFT exhibits a computational cost that scales linearly with the system size, irrespective of the system state $N$, while the Kohn-Sham DFT scales as $O(N^3)$.
This difference becomes particularly significant for large systems, especially the crust of neutron stars.
The stumbling block of orbital-free DFT is how to seek sufficiently accurate descriptions of kinetic energy with the density alone.
Currently, within the framework of non-relativistic DFT, the exact kinetic energy density functional is known for only two cases.
One case is for the one-particle system, which is the von Weizs{\"a}cker (vW) kinetic energy density functional (KEDF)~\cite{Weizsaecker1935Z.Physik},
\begin{equation}\label{vW}
  T_{\rm vW} = \frac{1}{2m}\int {\rm d}^3 r ~(\nabla \sqrt{\rho})^2 .
\end{equation}
Note that the vW kinetic functional~(\ref{vW}) can also be regarded as an exact KEDF for the ground states of many-boson systems.
Another case is for the uniform system, which is the Thomas-Fermi (TF) KEDF~\cite{Thomas1927, Fermi1927},
\begin{equation}\label{TF}
  T_{\rm TF} = \frac{1}{2m}\int {\rm d}^3 r ~\frac{3}{5}\left(3 \pi^2 \right)^{2/3} \rho^{5/3} .
\end{equation}
The TF functional is derived from the local implementation of a uniform gas model and is known to be exact when the number of electrons tends to infinity.
These two KEDFs as well as their combinations have been widely used in practical nuclear structure calculations~\cite{Brack1972Rev.Mod.Phys., Bohigas1976Phys.Lett.B, Brack1985Phys.Rep., Dutta1986Nucl.Phys.A, Aboussir1995Atom.DataNucl.DataTables, Centelles1990Nucl.Phys.A, Centelles2007Ann.Phys.},
and also serve as starting points to seek for more accurate KEDFs of non-relativistic DFT~\cite{Brack1973Nucl.Phys.A, Brack1985Phys.Rep., Centelles2007Ann.Phys., Wu2022Phys.Rev.C, Colo2023PTEP, Chen2024IJMPE, Wu2025CP, Wu2025arXiv}.

Relativistic density functional theory (RDFT) combines ideas of quantum field theory and density functional theory, which has gained wide attention for many attractive advantages, such as the automatic inclusion of the nucleonic spin degree of freedom and the spin-orbital interaction, the relativistic saturation mechanism, the isospin dependence of the spin-orbit potential, the consistent treatment of time-odd fields, the explanation of the pseudospin symmetry~\cite{Meng2016book, Ring2012PhysicaScripta}.
The RDFT has also been widely applied to nuclear physics within the KS scheme.
In the relativistic case, the wavefunction is a four-component spinor, which is much more complicated than the non-relativistic case.
The computational demands are thus relatively higher.
Therefore, the OF-DFT would be in principle quite attractive in the relativistic cases.
However, research on relativistic OF-DFT is still a blank in the field.

In this work, we take the first step toward relativistic orbital-free DFT. We derive the relativistic orbital-free kinetic energy density functional for a one-particle system in one-dimensional case, which can serve as a starting point for further exploration of more general relativistic orbital-free KEDFs.


\section{Derivation of kinetic energy density functional}

Here we present the derivation of the exact kinetic energy density functional for a one-particle system in both non-relativistic and relativistic cases.
For simplicity, we consider the one-dimensional case.


Firstly, we provide a brief review of the derivation of the exact kinetic energy density functional for one-particle systems in the non-relativistic case, based on the Schr{\"o}dinger equation~\cite{Weizsaecker1935Z.Physik}.

The static Schr{\"o}dinger equation for particles with mass $m$ in an external potential $V(x)$ reads $(\hbar=1)$
\begin{equation}\label{Sch}
  \left[-\frac{1}{2m}\frac{{\rm d}^2}{{\rm d}x^2}+V(x)\right]\psi_i(x) = E_i\psi_i(x).
\end{equation}
Density $\rho(x)$ can be calculated from wavefunctions $\psi_i(x)$,
\begin{equation}\label{density}
  \rho(x)=\sum_i|\psi_i(x)|^2.
\end{equation}
In the case of one-particle systems, the particle would occupy the orbital with lowest energy, simply denoted as $\psi(x)$.
The wave function corresponding to the first orbital does not have any sign-changing point, therefore its relation to density can be expressed as
\begin{equation}\label{wave}
  \psi(x)=\sqrt{\rho(x)}.
\end{equation}
Kinetic energy $T$ of the particle can be calculated via definition,
\begin{align}
  T[\rho] = & \int \psi(x)\left(-\frac{1}{2m}\frac{{\rm d}^2}{{\rm d}x^2}\right)\psi(x) {\rm d}x \notag \\
    = & \int \sqrt{\rho(x)}\left(-\frac{1}{2m}\frac{{\rm d}^2}{{\rm d}x^2}\right)\sqrt{\rho(x)} {\rm d}x  \notag \\
    = & \frac{1}{2m} \int \left(\frac{{\rm d}}{{\rm d}x}\sqrt{\rho(x)}\right)^2{\rm d}x  =  \frac{1}{8m}\int \frac{1}{\rho(x)}\left(\frac{{\rm d}\rho(x)}{{\rm d}x}\right)^2 {\rm d}x. \label{E_kin_1}
\end{align}
Its functional derivative $\frac{\delta T}{\delta\rho}$ is then obtained as
\begin{align}
  \frac{\delta T}{\delta\rho} = & \frac{1}{8m} \left[ \left(\frac{{\rm d}\rho}{{\rm d}x}\right)^2  \left(-\frac{1}{\rho^2}\right) + 2 \left(\frac{{\rm d}\rho}{{\rm d}x}\right)^2 \frac{1}{\rho^2} - 2\frac{1}{\rho} \frac{{\rm d}^2\rho}{{\rm d}x^2} \right] \notag \\
  =& \frac{1}{8m} \left[ \frac{1}{\rho^2}  \left(\frac{{\rm d}\rho}{{\rm d}x}\right)^2  - \frac{2}{\rho} \frac{{\rm d}^2\rho}{{\rm d}x^2} \right]
  = -\frac{1}{2m\sqrt{\rho(x)}}\frac{{\rm d}^2\sqrt{\rho(x)}}{{\rm d} x^2}. \label{T_rho_FD}
\end{align}

The case with multiple particles would be more complicated, which does not have an exact analytical expression yet.
However, such expression should be equivalent to Eq.~\eqref{E_kin_1} as the particle number approaching one.
Therefore, Eq.~\eqref{E_kin_1} can provide a benchmark for the study of more complex multi-particle cases.


Next, we present a detailed derivation of the exact kinetic energy density functional and its functional derivatives for one-particle nuclear systems in the relativistic case, based on the Dirac equation.


The one-dimensional Dirac equation for particles with mass $m$ under vector potential $V(x)$ and scalar potential $S(x)$ reads
\begin{equation}\label{Dirac_eq}
  \left\{-i\alpha\partial_x + V(x) + \beta [m+S(x)]\right\}\Psi_i(x) = E_i\Psi_i(x),
\end{equation}
where
\begin{equation}\label{matrix_alp_bet}
  \alpha = \begin{pmatrix} 0&1\\1&0 \end{pmatrix},~~~~~~~~\beta = \begin{pmatrix} 1&0\\0&-1 \end{pmatrix}.
\end{equation}
The solutions of Eq.~\eqref{Dirac_eq} are binary wavefunctions, which are denoted as $\Psi_i(x)=\begin{pmatrix} \psi_{1i}(x)\\ i\psi_{2i}(x) \end{pmatrix}$.

The vector density and scalar density are defined respectively as
\begin{align}
  \rho_v(x) = & \sum_i \overline{\Psi}_i(x) \beta \Psi_i(x), \\
  \rho_s(x) = & \sum_i \overline{\Psi}_i(x) \Psi_i(x).
\end{align}
In the one-particle case, they read
\begin{align}
  \rho_v(x) = & \overline{\Psi}(x) \beta \Psi(x) = \psi_1^2(x) + \psi_2^2(x) \label{Eqrhov} \\
  \rho_s(x) = & \overline{\Psi}(x) \Psi(x) = \psi_1^2(x) - \psi_2^2(x),\label{Eqrhos}
\end{align}
where $\Psi(x)=\begin{pmatrix} \psi_{1}(x)\\ i\psi_{2}(x) \end{pmatrix}$ denotes the ground-state wavefunction.

The kinetic energy density $\tau(x)$ is calculated as
\begin{align}
  \tau(x) = & \Psi^+(x) (-i\alpha\partial_x + \beta m) \Psi(x) \notag \\
    = & \begin{pmatrix} \psi_1(x)& -i\psi_2(x) \end{pmatrix} \begin{pmatrix} m&-i\partial_x\\-i\partial_x&-m \end{pmatrix} \begin{pmatrix} \psi_1(x)\\ i\psi_2(x) \end{pmatrix} \notag \\
    = & m\rho_s(x) +\psi_1(x)\psi_2'(x) - \psi_2(x)\psi_1'(x). \label{tden1}
\end{align}
Kinetic energy $T$ is calculated by integrating kinetic energy density $\tau(x)$.
In the sense of integration, one has
\begin{equation}
  \int \psi_1(x)\psi_2'(x) {\rm d}x + \int \psi_1'(x)\psi_2(x) {\rm d}x = \left.\psi_1(x)\psi_2(x)\right|_{-\infty}^{+\infty} = 0,
\end{equation}
and, thus, for simplification, the kinetic energy density can be written as
\begin{align}
  \tau(x) = & m\rho_s(x)  - 2\psi_2(x)\psi_1'(x) \label{tau1} \\
  = & m\rho_s(x) + 2\psi_1(x)\psi_2'(x). \label{tau2}
\end{align}

From Eqs.~\eqref{Eqrhov} and \eqref{Eqrhos}, one can obtain
\begin{align}
    \psi_1^2(x) = & \frac{\rho_v(x)+\rho_s(x)}{2}, \\
    \psi_2^2(x) = & \frac{\rho_v(x)-\rho_s(x)}{2}.
\end{align}
For convenience, the following notations are defined,
\begin{align}
   \rho_{+}(x) = & \rho_v(x) + \rho_s(x), \\
   \rho_{-}(x) = & \rho_v(x) - \rho_s(x).
\end{align}
For the lowest orbital, $\psi_1(x)$ does not have sign-changing points, while $\psi_2(x)$ has a sign-changing point.
Therefore, the wave functions can be expressed as
\begin{align}
    \psi_1(x) = & \sqrt{\frac{\rho_{+}(x)}{2}}, \label{Eqpsi1} \\
    \psi_2(x) = & -S_{\rm ign}\sqrt{\frac{\rho_{-}(x)}{2}}, \label{Eqpsi2}
\end{align}
where $|S_{\rm ign}|=1$ and changes its sign crossing the sign-changing point.
At the sign-changing point, as can be seen in Eq.~\eqref{Eqpsi2}, $\psi_2(x) = \sqrt{\frac{\rho_{-}(x)}{2}} =0$.
Therefore, $S_{\rm ign}$ can be defined as any finite value at that point, e.g., $S_{\rm ign}=1$.
We will ultimately validate the equations obtained at the sign-changing point after the following derivation.
With Eqs.~\eqref{Eqpsi1} and \eqref{Eqpsi2}, the kinetic energy density can thus be expressed as
\begin{align}
   \tau(x) = & m\rho_s(x) - 2\psi_2(x)\psi_1'(x) \notag \\
        = & m\rho_s(x) + \frac{S_{\rm ign}}{2}\sqrt{\frac{\rho_{-}(x)}{\rho_{+}(x)}}[\rho_{+}'(x)]. \label{tau_eq1}
\end{align}

The only issue remaining in expressing kinetic energy $\tau(x)$ with densities $(\rho_v, \rho_s)$ is to determine the value of $S_{\rm ign}$.
From the Dirac equation \eqref{Dirac_eq}, one can obtain
\begin{align}
  [E-V(x)-m-S(x)]\psi_1(x) = & \psi_2'(x), \\
  [V(x)-m-S(x)-E]\psi_2(x) = & \psi_1'(x). \label{Dirac_mod}
\end{align}
For real nuclei, one has $[V(x)-m-S(x)-E]<0$, and thus, $\psi_2(x)$ has a different sign from $\psi_1'(x)$ due to Eq.~\eqref{Dirac_mod}.
This gives
\begin{equation} \label{Sign_psi}
   -S_{\rm ign}=-\frac{|\psi_1'(x)|}{\psi_1'(x)},
\end{equation}
when $\psi'_1(x) \neq 0$.
It corresponds to the sign-changing point when $\psi'_1(x) = 0$.
The $\psi_1'(x)$ can be calculated from $\rho_v$ and $\rho_s$ as
\begin{equation}
   \psi_1'(x)=\left(\sqrt{\frac{\rho_{+}(x)}{2}}\right)'=\frac{1}{2}\sqrt{\frac{2}{\rho_{+}(x)}}\frac{\rho_{+}'(x)}{2}.
\end{equation}
Therefore, the $S_{\rm ign}$ can be expressed as
\begin{equation}
  S_{\rm ign} = \frac{\left|\rho_{+}'(x)\right|}{\rho_{+}'(x)}, \label{Sign_Eq}
\end{equation}
when $\rho_v'(x)+\rho_s'(x) \neq 0$.
It corresponds to the sign-changing point when $\rho_v'(x)+\rho_s'(x) = 0$.
By substituting $S_{\rm ign}$ given by Eq.~\eqref{Sign_Eq} into Eq.~\eqref{tau_eq1}, one obtains
\begin{equation}
  \tau(x) =  m\rho_s(x) + \frac{1}{2}\sqrt{\frac{\rho_{-}(x)}{\rho_{+}(x)}}\left|\rho_{+}'(x)\right| .
\end{equation}
Therefore, the kinetic energy density functional $T[\rho_v(x),\rho_s(x)]$ can now be expressed as
\begin{align}
  &T[\rho_v(x), \rho_s(x)] \notag \\
  = &  m\int\rho_s(x){\rm d}x + \int \frac{S_{\rm ign}}{2}\sqrt{\frac{\rho_{-}(x)}{\rho_{+}(x)}}\left[\rho_{+}'(x)\right] {\rm d}x \notag  \\
  = & m\int\rho_s(x){\rm d}x  + \frac{1}{2} \int  \sqrt{\frac{\rho_{-}(x)}{\rho_{+}(x)}}\left|\rho_{+}'(x)\right| {\rm d}x \label{T_rhovs1}.
\end{align}
We have now obtained the exact kinetic energy density functional for one-particle nuclear systems in the relativistic case.
The next step is to derive its functional derivatives.

Usually, the mass term should be deducted from the relativistic kinetic energy, and  Eq.~\eqref{T_rhovs1} is rewritten as
\begin{align}
  &T[\rho_v(x), \rho_s(x)] \notag \\
  = &  m\int\rho_s(x){\rm d}x - m\int\rho_v(x){\rm d}x + \int\frac{S_{\rm ign}}{2}\sqrt{\frac{\rho_{-}(x)}{\rho_{+}(x)}}\left[\rho_{+}'(x)\right] {\rm d}x \notag \\
  = & -m\int\rho_{-}(x){\rm d}x + \int\frac{S_{\rm ign}}{2}\sqrt{\frac{\rho_{-}(x)}{\rho_{+}(x)}}\left[\rho_{+}'(x)\right] {\rm d}x, \label{Tkin_mf}
\end{align}
where $S_{\rm ign}$ is kept for simplification.

The functional derivatives $\frac{\delta T}{\delta\rho_{+}}$ and $\frac{\delta T}{\delta\rho_{-}}$ can be thus derived as
\begin{align}
  \frac{\delta T}{\delta\rho_{+}} = & \frac{S_{\rm ign}}{2} \frac{\partial}{\partial \rho_{+}}\left[(\rho_{-})^{1/2}(\rho_{+})^{-1/2}\rho'_{+}\right] \notag \\
  &- \frac{S_{\rm ign}}{2}\frac{{\rm d}}{{\rm d}x}\left\{\frac{\partial}{\partial\rho'_{+}} \left[(\rho_{-})^{1/2}(\rho_{+})^{-1/2}\rho'_{+}\right]  \right\} \notag \\
  = & - \frac{S_{\rm ign}}{4}(\rho_{-})^{-1/2}\rho'_{-}(\rho_{+})^{-1/2}, \label{Eq_T_vps2}
\end{align}
and
\begin{align}
  \frac{\delta T}{\delta\rho_{-}} = & -m + \frac{S_{\rm ign}}{2} \frac{\partial}{\partial \rho_{-}}\left[(\rho_{-})^{1/2}(\rho_{+})^{-1/2}\rho'_{+}\right] \notag \\
  =& -m + \frac{S_{\rm ign}}{4} (\rho_{-})^{-1/2}(\rho_{+})^{-1/2}\rho'_{+}, \label{Eq_T_vms2}
\end{align}
respectively.

Up to this point, we have derived the exact kinetic energy density functional \eqref{Tkin_mf} and its functional derivatives \eqref{Eq_T_vps2} and \eqref{Eq_T_vms2} for one-particle nuclear systems in the relativistic DFT.

The sign of $S_{\rm ign}$ changes at the point $x_{\rm node}$ where $\rho'_{+}=0$ and $\rho_{-}=0$, as can be seen from Eqs.~\eqref{Dirac_mod}, \eqref{Sign_psi}, and \eqref{Sign_Eq}.
Therefore, the integrand in Eq.~\eqref{Tkin_mf} for kinetic energy $T[\rho_{+}, \rho_{-}]$ equals $0$ at that point, which does not give rise to a divergence problem.
However, $\rho_{-}=0$ might lead to a singularity in Eqs.~\eqref{Eq_T_vps2} and \eqref{Eq_T_vms2} because of $(\rho_{-})^{-1/2}$.
Meanwhile, $\rho'_{-}$ and $\rho'_{+}$ also equal to $0$ at the sign-changing point, which means that $\frac{0}{0}$ structures appear in both Eqs.~\eqref{Eq_T_vps2} and \eqref{Eq_T_vms2}.
Therefore, one should carefully examine the limit values at the sign-changing point.

The Taylor expansions of $\rho_{-}$, $\rho'_{-}$, and $\rho'_{+}$ around the sign-changing point $x_{\rm node}$ are as follows,
\begin{align}
  \rho_{-}(x) = & \underbrace{\rho_{-}(x_{\rm node})}_{0} +  \underbrace{\rho'_{-}(x_{\rm node})}_{0}\cdot(x-x_{\rm node}) \notag \\
  & + \frac{1}{2!} \rho''_{-}(x_{\rm node})\cdot(x-x_{\rm node})^2 + ... \label{limitvms} \\
  \rho'_{-}(x) = & \underbrace{\rho'_{-}(x_{\rm node})}_{0} + \rho''_{-}(x_{\rm node})\cdot(x-x_{\rm node})+ ... \label{limitvmsp}\\
  \rho'_{+}(x) = & \underbrace{\rho'_{+}(x_{\rm node})}_{0} + \rho''_{+}(x_{\rm node})\cdot(x-x_{\rm node})+ ...  \label{limitvpsp}
\end{align}
With Eqs.~\eqref{limitvms}, \eqref{limitvmsp}, and \eqref{limitvpsp}, up to the first non-zero term, one has
\begin{align}
  (\rho_{-})^{-1/2}\rho'_{-} = & \frac{\rho''_{-}(x_{\rm node})\cdot(x-x_{\rm node})}{\sqrt{\frac{1}{2!} \rho''_{-}(x_{\rm node})\cdot(x-x_{\rm node})^2}} \notag \\
   = &  \sqrt{2}\sqrt{\rho''_{-}(x_{\rm node})} \cdot\frac{(x-x_{\rm node})}{|x-x_{\rm node}|}. \\
  (\rho_{-})^{-1/2}\rho'_{+} = & \frac{\rho''_{+}(x_{\rm node})\cdot(x-x_{\rm node})}{\sqrt{\frac{1}{2!} \rho''_{-}(x_{\rm node})\cdot(x-x_{\rm node})^2}} \notag \\
  & =  \sqrt{2}\frac{ \rho''_{+}(x_{\rm node})}{\sqrt{\rho''_{-}(x_{\rm node})}} \cdot\frac{(x-x_{\rm node})}{|x-x_{\rm node}|}.
\end{align}
Note that $\rho_{+}$ increases when $x<x_{\rm node}$, and decreases when $x>x_{\rm node}$.
Therefore, $S_{\rm ign}=|\rho'_{+}|/\rho'_{+}$ is positive when $x<x_{\rm node}$ and negative when $x>x_{\rm node}$.
As a result, $-S_{\rm ign}\frac{(x-x_{\rm node})}{|x-x_{\rm node}|}=1$ both before and after the sign-changing point.
In the neighborhood of the sign-changing point, the functional derivatives can now be calculated as
\begin{align}
  \frac{\delta T}{\delta\rho_{+}} = & \frac{\sqrt{2}}{4} (\rho_{+})^{-1/2}\sqrt{\rho''_{-}},  \label{fd_nei1} \\
  \frac{\delta T}{\delta\rho_{-}} = & -m - \frac{\sqrt{2}}{4}(\rho_{+})^{-1/2}\frac{\rho''_{+}}{\sqrt{\rho''_{-}}}. \label{fd_nei2}
\end{align}
Note that the values involved in Eqs.~\eqref{fd_nei1} and \eqref{fd_nei2} are now all finite ones.


\section{Verification of the functional}

The subsequent step is to verify the obtained kinetic energy density functional and its functional derivatives.
We consider a system consisting of non-interacting particles trapped in the relativistic potentials.
The energy of relativistic system can be written as a functional of the vector density $\rho_v$ and the scalar density $\rho_s$,
\begin{equation}\label{Eq_Etot_pot}
  E_{\rm tot}[\rho_{v}, \rho_{s}] =  T[\rho_{v},\rho_{s}] + E_{\rm {pot.}}[\rho_{v},\rho_{s}],
\end{equation}
where the potential energy associated with the vector potential $V(x)$ and the scalar potential $S(x)$ is
\begin{align}
  E_{\rm {pot.}}[\rho_{v},\rho_{s}] = & \int {\rm d}x (V\rho_v + S\rho_s) \notag \\
  = & \int {\rm d}x \left( \frac{V+S}{2}\rho_{+} + \frac{V-S}{2}\rho_{-} \right). \label{Eq_Etot}
\end{align}
The self-consistent solution of this density functional can be obtained by varying the total energy with respect to the densities $(\rho_{v}, \rho_{s})$, under the  constraint of particle number conservation,
\begin{equation}\label{Eq_var}
  \delta\left\{T[\rho_{+},\rho_{-}] + E_{\rm {pot.}}[\rho_{+},\rho_{-}]
 - \frac{\mu}{2}\int {\rm d}x[\rho_{+} + \rho_{-}] \right\} = 0,
\end{equation}
where $\mu$ is adjusted to produce the required particle number.
The detailed version of Eq.~\eqref{Eq_var} is as follows,
\begin{align}\label{Eq_var1}
  & \int \delta\rho_{+}\left\{\frac{\delta T[\rho_{+},\rho_{-}]}{\delta\rho_{+}} + \frac{\delta E_{{\rm pot.}}[\rho_{+},\rho_{-}]}{\delta\rho_{+}} -\frac{\mu}{2}\right\} {\rm d} x  \notag \\
  & + \int \delta\rho_{-}\left\{\frac{\delta T[\rho_{+},\rho_{-}]}{\delta\rho_{-}} + \frac{\delta E_{{\rm pot.}}[\rho_{+},\rho_{-}]}{\delta\rho_{-}} -\frac{\mu}{2}\right\} {\rm d} x = 0.
\end{align}
Since Eq.~\eqref{Eq_var1} should hold for all $\delta\rho_{+}$ and $\delta\rho_{-}$, one has
\begin{align}
  & \frac{\delta T[\rho_{ +},\rho_{ -}]}{\delta\rho_{+}} = - \frac{V+S}{2} +\frac{\mu}{2}, \label{FD_V}\\
  & \frac{\delta T[\rho_{+},\rho_{-}]}{\delta\rho_{-}} = - \frac{V-S}{2} +\frac{\mu}{2}. \label{FD_S}
\end{align}
Note that Eqs.~\eqref{Eq_Etot_pot} and \eqref{Eq_Etot} for energy density functional and Eqs.~\eqref{FD_V} and \eqref{FD_S} for functional derivatives are not only applicable to a one-particle system but also to multi-particle systems.

In order to verify kinetic energy and its functional derivatives through Eqs.~\eqref{FD_V} and \eqref{FD_S}, we numerically solve the Dirac equation~\eqref{Dirac_eq} with a one-particle system under the Woods-Saxon types of potentials,
\begin{align}
  V+S = & \frac{U_0}{1+\exp[(|x|-x_0)/a_0]}, \label{Eq_VplusS} \\
  V-S = & -\lambda\frac{U_0}{1+\exp[(|x|-x_0)/a_0]}. \label{Eq_VminusS}
\end{align}
The parameters in Eqs.~\eqref{Eq_VplusS} and \eqref{Eq_VminusS} are taken as $U_0 = -67~{\rm MeV}$, $x_0 = 5.5~{\rm fm}$, $a = 0.6~{\rm fm}$, and $\lambda = 11$.
These values are matched with typical cases for nuclear physics.

The Dirac equation~\eqref{Dirac_eq} is solved with the shooting method.
After solving the Dirac equation, one obtains the wavefunction $\Psi(x)=\begin{pmatrix} \psi_1(x) \\ i\psi_2(x) \end{pmatrix}$.
With the wavefunction, one can calculate the kinetic energy for the one-particle system
\begin{equation}\label{Eq_Ekin_numeral}
  T^{\rm wavefunction} = \int \Psi^+(x) (-i\alpha\partial_x + \beta m) \Psi(x)  {\rm d}x - m,
\end{equation}
and also the densities via Eqs.~\eqref{Eqrhov} and \eqref{Eqrhos}.

The kinetic energy calculated with Eq.~\eqref{Eq_Ekin_numeral} is $T^{\rm wavefunction}=0.775780~{\rm MeV}$.
Once the densities $\rho_{+}$ and $\rho_{-}$ are obtained, one can also calculate the kinetic energy $T^{\rm functional}$ with Eq.~\eqref{Tkin_mf}, which is $T^{\rm functional}=0.775773~{\rm MeV}$.
They are consistent within an acceptable numerical error.

\begin{figure}[!h]
\centering
\includegraphics[width=\linewidth]{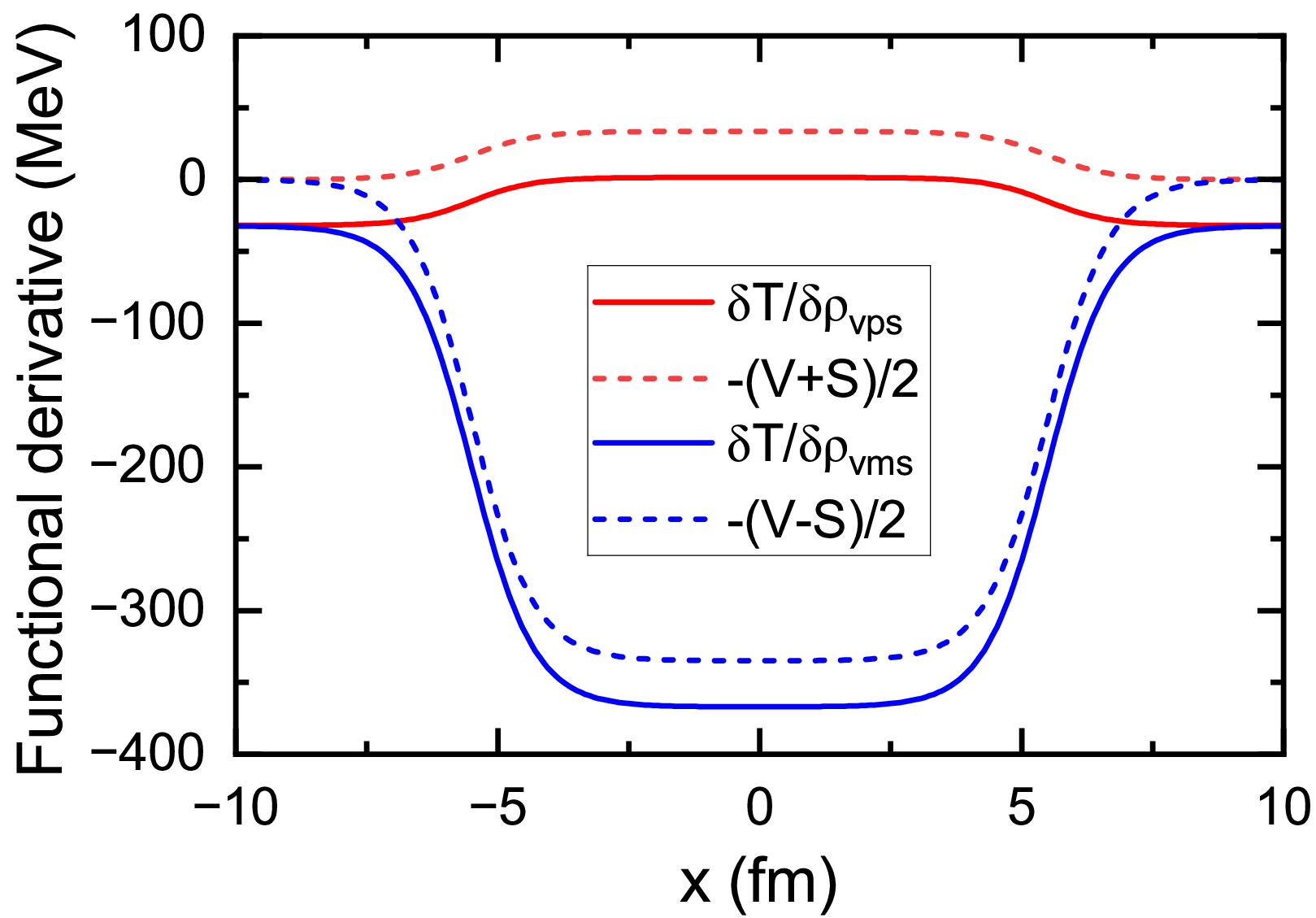}
\caption{
Verifications of the functional derivatives~\eqref{Eq_T_vps2} and \eqref{Eq_T_vms2} through Eqs.~\eqref{FD_V} and \eqref{FD_S}.
}
\label{fig1}
\end{figure}

One can also calculate the functional derivatives $\frac{\delta T}{\delta\rho_{+}}$ and $\frac{\delta T}{\delta\rho_{-}}$ through Eqs.~\eqref{Eq_T_vps2} and \eqref{Eq_T_vms2}.
Equations~\eqref{FD_V} and \eqref{FD_S} can then be used to validate the correctness of functional derivatives.
They are also consistent within acceptable numerical errors.
As can be seen in Fig.~\ref{fig1}, the curve of $\frac{\delta T}{\delta\rho_{+}}$ and the curve of $-(V+S)/2$ can be matched by translation with a constant related to $\mu/2$.
The same translation holds for the curve of $\frac{\delta T}{\delta\rho_{-}}$ and the curve of $-(V-S)/2$.
This means that Eqs.~\eqref{FD_V} and \eqref{FD_S} are fully fulfilled and the functional derivatives \eqref{Eq_T_vps2} and \eqref{Eq_T_vms2} are verified to be correct.
Note that functional derivatives at the sign-changing point, which is $x_{\rm node}=0$ in the current validation, are calculated with Eqs.~\eqref{fd_nei1} and \eqref{fd_nei2}.
As can be seen in Fig.~\ref{fig1}, the values at this point continuously match the full functions.

Up to this point, we have verified the correctness of newly derived exact kinetic energy density functional \eqref{Tkin_mf} and its functional derivatives \eqref{Eq_T_vps2} and \eqref{Eq_T_vms2} for one-particle nuclear systems in the relativistic case.


\section{Summary}

The exact relativistic orbital-free kinetic energy density functional for one-particle nuclear systems and its functional derivatives have been derived in the one-dimensional case.
Both the derived kinetic energy density functional and its functional derivatives have been verified.
This provides a very important first step toward nuclear relativistic orbital-free DFT.
The derived formulas can serve as starting points for further exploration of more general relativistic orbital-free KEDFs.



{\it Acknowledgments.} This work was supported by the National Natural Science Foundation of China (Grants No. 12405134, No. 12435006, No. 12141501, No. 12475117, No. 11935003, No. 12141501), the National Key R\&D Program of China (Contract No. 2024YFE0109803), the State Key Laboratory of Nuclear Physics and Technology, Peking University (Grant No. NPT2023ZX03), the National Key Research and Development Program of China 2024YFA1612600, and the National Key Laboratory of Neutron Science and Technology (Grant No. NST202401016).
Z.X.R. is supported in part by the European Research Council (ERC) under the European Union's Horizon 2020 research and innovation programme (Grant Agreement No. 101018170).

{\it Data availability.} The dataset can be accessed upon request to the corresponding author.

\end{document}